# Electrochemical kinetics of SEI growth on carbon black

# I: Experiments

Peter M. Attia[1], Supratim Das[2], Stephen J. Harris[3], Martin Z. Bazant[2], William C. Chueh[1*]

*Corresponding author: wchueh@stanford.edu

1. Department of Materials Science and Engineering, Stanford University, Stanford, CA 94305
2. Department of Chemical Engineering, Massachusetts Institute of Technology, Cambridge, MA 02139
3. Materials Science Division, Lawrence Berkeley National Laboratory, Berkeley, CA 94720

## Abstract

Growth of the solid electrolyte interphase (SEI) is a primary driver of capacity fade in lithium-ion batteries. Despite its importance to this device and intense research interest, the fundamental mechanisms underpinning SEI growth remain unclear. In Part I of this work, we present an electroanalytical method to measure the dependence of SEI growth on potential, current magnitude, and current direction during galvanostatic cycling of carbon black/Li half cells. We find that SEI growth strongly depends on all three parameters; most notably, we find SEI growth rates increase with nominal C rate and are significantly higher on lithiation than on delithiation. We observe this directional effect in both galvanostatic and potentiostatic experiments and discuss hypotheses that could explain this observation. This work identifies a strong coupling between SEI growth and charge storage (e.g., intercalation and capacitance) in carbon negative electrodes.

Improving the lifetime of lithium-ion batteries will enable emerging technologies like electric vehicles and lower the considerable environmental impact of disposed batteries. A dominant aging mechanism in lithium-ion batteries is growth of the solid-electrolyte interphase (SEI), the product of continuous electrolyte reduction at low potentials on the surface of the graphitic negative electrode.[1–6] A deeper understanding of SEI growth – specifically, its dependence on electrochemical cycling conditions – could unlock dramatic improvements in lithium-ion battery lifetime.

While extensive physical characterization of the SEI on graphite provides insight into its composition, morphology, and chemistry[6–16], a major challenge in SEI characterization is its thin and pseudo-amorphous nature and its sensitivity to air[4,17,18] and x-ray/electron radiation[19–21]. Electrochemical characterization provides a quantitative and *in situ* method to measure SEI growth, specifically as a function of cycling conditions. In particular, coulometry has emerged as an electrochemical technique to precisely capture the irreversible capacity loss within a cycle.[22–26] Smith *et al.*[26] utilized high-precision coulometry to measure the cumulative sum of the difference between charge and discharge capacities during cycling in a graphite/lithium half cell, attributing this irreversible capacity loss to SEI growth. Complementary electrochemical techniques to coulometry include differential capacity analysis[27–30] (*dQ/dV*) and delta differential capacity analysis[31], (Δ*dQ/dV*) where Δ specifies differences between cycles. With these methods, transformations of voltage-capacity curves provide insight into specific degradation modes such as loss of lithium inventory, loss of active material, and impedance rise.

Previous work has revealed the dependencies of graphitic SEI growth on key electrochemical parameters, including time, electrode potential, and C rate. The time dependence of SEI growth is well studied: experimental measurements have captured its linear relationship with the square root of time, $t^{0.5}$, by measuring the capacity decrease[32–36] or impedance increase[37,38] during battery storage[33,34,37,38] and cycling[32,35–37]. Consequently, SEI growth is often modeled as mixed reaction/diffusion-limited film growth[39–44], approaching $t^{0.5}$ scaling in the long time limit. Most work assumes the SEI reaches this transport-limited regime quickly, i.e., by the end of the first cycle. Additionally, a few studies have shown that the SEI growth rate in full cells increases with cell voltage (corresponding to decreased negative electrode potential) during storage[45] and cycling[46]. These studies correlate the full-cell potential with capacity fade by discretizing cutoff or storage potentials and sampling over many cells. Lastly, C rate was shown via coulometry[35] and modeling[43] to have a weak effect on SEI growth rate, at least for low (<C/10) nominal rates. We can qualitatively summarize these dependencies of SEI growth as "time spent at low potential".

Generally, SEI growth has been considered to be analogous to self-passivating oxide growth on metals and semiconductors[47] due to their shared $t^{0.5}$ dependence.[39–44] However, SEI growth and conventional passivation layer growth have two important differences. First, SEI can grow via both chemical and electrochemical reactions. Chemical SEI growth, in which the source of lithium ions is the lithiated electrode, is expected during storage (affecting "calendar life") and cycling (affecting "cycle life"), while electrochemical SEI growth, in which the source of lithium ions is the electrolyte, is expected only during cycling. These processes have been treated similarly in the literature despite their significant mechanistic differences. Second, the "substrate" for SEI growth is electrochemically active during battery operation – that is, the electrode is dynamically storing and releasing charge. Because lithium ions from the electrolyte and electrons from the electrode are shared reactants in both charge storage modes (e.g., intercalation and capacitance) and SEI growth, these processes likely do not occur independently. In other words, a coupling between reversible charge storage and irreversible SEI growth is expected. If this coupling between SEI growth and intercalative charge storage were to exist, the mechanism of SEI growth on "inert" electrodes like copper would differ significantly from the growth mechanism on intercalating electrodes like carbon or silicon, even if the SEI chemistries and morphologies were similar.

An interesting aspect of this possible coupling is the difference in SEI growth rate between carbon lithiation and delithiation, within the same cycle, for a given potential. In this work, we use the terms lithiation and delithiation to refer to both intercalation and capacitive charge storage. To the best of the authors' knowledge, this dependence of SEI growth on current direction has not been studied, as previous work integrated SEI growth across a charge/discharge cycle and does not capture intracycle differences. Here, we provide two clarifications. First, this question of "directional symmetry" in SEI growth as a function of carbon (de)lithiation direction at the same potential is distinct from that of the difference between SEI growth at low potential and SEI decomposition at high potential, which has been previously studied.[48–55] Second, we distinguish first cycle SEI growth from "post-first-cycle" growth; the first cycle SEI reaction is confined specifically to the first lithiation of carbon, as evidenced by electrochemical and gas evolution studies.[56–59] Post-first-cycle SEI growth, the more relevant regime in lithium-ion battery degradation, is the focus of this study.

As illustrated in Figure 1, a simple metric of degree of directional symmetry in SEI growth between lithiation and delithiation of carbon is capacity vs cycle number. The measured capacity is the sum of the reversible electrode capacity (dotted red line) and the SEI capacities, which decay with cycle number due to the SEI's self-passivating nature. Three limiting cases of directional symmetry in SEI growth are possible (assuming SEI growth is irreversible): SEI only grows during carbon lithiation (1a), SEI grows in both directions (1b), or SEI only grows during carbon delithiation (1c). We term 1a and 1c as asymmetric growth and 1b as symmetric growth. The irreversible SEI capacity adds to the reversible capacity of the carbon electrode during lithiation (1d) and subtracts from this capacity during delithiation (1e). Quantifying and understanding this dependence would reveal mechanistic insight into SEI growth, specifically its coupling to intercalative and capacitive charge storage, and aid modeling of battery degradation.

In the absence of a coupling mechanism as described above, a consideration of the thermodynamic driving forces for this process would indicate that SEI growth should occur during both lithiation and delithiation (1b). Neglecting the Nernstian concentration term, the driving force for the SEI reaction can be approximated by $\eta = E_0 - E$, where $E_0$ is the standard redox potential of SEI growth (canonically, 0.8 V vs Li$^+$/Li$^0$ for 1.0 M LiPF$_6$ in 50:50 wt. % EC:DEC electrolyte or similar[4,17,60]) and $E$ is the electrode potential. Since the potential profiles of a slowly-cycled electrode without SEI are nearly overlapping for both lithiation and delithiation (particularly for non-phase-separating materials), the driving force for SEI growth is also similar in both directions. Accordingly, roughly symmetric growth (1b) is expected if SEI growth is a function of only time and potential, as the literature would suggest.

Most previous SEI work is performed on graphite, the most commercially relevant negative electrode material today. However, electrochemical SEI characterization on graphite is challenging due to its low specific surface area; in other words, the "signal" of irreversible SEI capacity is dwarfed by the "background" of reversible graphitic capacity. Additionally, the multiple phases present in graphite introduce additional complexity into its role as a substrate for SEI growth. Other carbon materials – specifically, nanomaterials like carbon black – grow significantly more SEI per cycle per unit mass and thus have a higher "signal-to-noise" ratio. Furthermore, its solid-solution lithiation pathway due to its disordered nanostructure allows us to study SEI growth without concern for the multiple phases and phase transitions present in graphite. Carbon black has graphitic ($sp^2$) carbon bonding and nanostructure, but its graphitic (crystalline) ordering extends no further than the nanometer length scale.[61–64] Previously, SEI growth on carbon black has been studied at low[15,26,48,65–68] and high[68,69] potentials, particularly in the context of its widespread use as a conductive electrode additive. Notably, Smith *et al.*[35] studied the SEI growth rate for electrodes with various compositions of graphite and carbon black and concluded that SEI growth occurs on the graphite and carbon black surfaces at approximately the same current densities (similar areal SEI growth rates). Thus, carbon black serves as a model carbon system with a high ratio of irreversible SEI capacity to reversible carbon capacity relative to graphite, as well as solid-solution lithiation. Importantly, our choice of an intercalating electrode over a non-intercalating electrode like copper enables us to study the coupling between SEI growth and bulk charge storage.

In Part I of this work, we experimentally investigate the electrochemical kinetics of SEI growth on carbon negative electrodes during (de)lithiation. We first establish the physical and electrochemical properties of carbon black. We then systematically examine SEI growth during the second cycle for carbon black/Li half cells under different galvanostatic cycling conditions, using delta differential capacity analysis to isolate SEI growth from reversible charge storage. Next, we evaluate the dependencies of SEI growth on potential, current magnitude, and current direction. Notably, we find a strong rate dependence and a stark directional asymmetry between SEI growth during lithiation and delithiation of carbon. We then perform additional experiments to confirm this result, concluding that a strong coupling between SEI growth and charge storage modes (intercalation and/or capacitance) exists, and discuss its possible origins. In Part II of this work, we investigate these hypotheses in depth and develop experimentally-validated models that explain these observations. Our results provide new insights into the electrochemistry of SEI growth on carbon negative electrodes, which can be used to reduce capacity fade in lithium-ion batteries.

**Experimental**

*Physical characterization.–* TIMCAL Super P carbon black nanopowder (Alfa Aesar) was used throughout this work, and Hitachi surface-modified graphite (SMG) battery-grade powder was used as a graphite reference sample for comparison to carbon black. Scanning electron microscopy (SEM) micrographs of carbon black electrode sheets were obtained with a FEI Magellan 400 XHR microscope, and transmission electron microscopy (TEM) micrographs of carbon black particles were obtained with a FEI Tecnai G2 F20 X-TWIN microscope. The bulk structure was characterized via X-ray diffraction (XRD) with a Bruker D8 diffractometer. Pair distribution function (PDF) measurements were performed at beamline 11-ID-B at the Advanced Photon Source at Argonne National Laboratory with an X-ray energy of 58.6 keV for pristine carbon black powder loaded into a 1 mm capillary tube. The pair distribution function *G(r)* was obtained by a Fourier transformation of the raw data using the PDFgetX2 software package. The carbon binding environments were characterized with X-ray photoelectron spectroscopy (XPS) on an PHI VersaProbe 3 instrument; in this experiment, the powders were dropcast onto copper foil and calibrated to the peaks of $Ar^+$-sputtered Cu $2p_{3/2}$.

*Electrode fabrication.–* Carbon black slurries were created with TIMCAL Super P and polyvinylidene difluoride (PVDF) binder (Alfa Aesar) in a 90:10 wt.% ratio with NMP solvent (Sigma-Aldrich). Slurries were mixed with a planetary mixer (THINKY AR-100), cast at a nominal thickness of 100 μm on electrodeposited copper foil (Hohsen), and dried overnight in a vacuum oven at 55 °C. The final electrode thickness ranged from 45–55 μm, as measured with a micrometer.

Electrode disks (13 mm diameter, 1.33 cm$^2$ geometric area) were then punched for coin cell assembly and weighed on an analytical microbalance (Mettler-Toledo XPR2). The active carbon black mass per disk is approximately 1 mg. The cells were assembled in an argon glove box (VAC, <1 ppm O$_2$ and <0.5 ppm H$_2$O) using stainless steel 2032 coin cell cases (Hoshen) with 50 μL of 1.0 M LiPF$_6$ in EC:DEC (1:1) by weight (BASF/Gotion Selectilyte LP40), one 25 μm separator (Celgard) and a lithium foil (Alfa Aesar) counter electrode. The geometric volume of the electrode is ~7 μL (1.33 cm$^2$ × 50 μm).

***Electrochemical characterization.–*** All cells were cycled inside a temperature chamber (AMEREX IC-150R) at a constant nominal temperature of 30.0 °C (±0.5 °C) with a Bio-logic BCS-805. Unless otherwise specified, all cells were charged and discharged at a constant current between 2.0 and 0.01 V, with no potentiostatic hold at either cutoff potential. All cells rest for 24 hours before cycling to ensure complete electrolyte wetting. To avoid convolution from first cycle SEI growth, each cell was subjected to one formation cycle at C/10, with a lower cutoff voltage of 0.01 V and an upper cutoff voltage of 2.0 V. All C rates were calculated with a nominal carbon black specific capacity of 200 mAh/g (1C = 200 mA/g$_{CB}$); these rates are referred to as nominal C rates. Per IUPAC convention, we consider discharging (carbon black lithiation) to be negative current and charging (carbon black delithiation) to be positive current.

## Results and Discussion

*Physical and electrochemical characterization of carbon black.–* Physical characterization of carbon black is presented in Figure 2. Figure 2A displays a SEM micrograph of a carbon black electrode, and Figure 2B displays a TEM micrograph of a single carbon black particle. In Figure 2B, the nanoscale graphitic domains within a single carbon black particle are clearly visible. Figures 2C and 2D display the XRD patterns of pristine graphite and carbon black powders. From the full diffraction patterns (Figure 2C), many 3D reflections are visible in graphite, while only 1D or 2D reflections are visible in carbon black. Pseudo-amorphous carbon materials like carbon black have turbostratic disorder, in which the basal plane layers in a stack have non-uniform interplanar spacing and random orientation.[61,62,70,71] This disorder prevents 3D reflections in diffraction patterns. The peak location of the (002) reflection for both materials (Figure 2D) in Super P is 5.2% larger than that of graphite. Deconvoluting the effects of crystallite size and strain for the large width of the (002) peak is difficult due to the lack of other peaks in the same crystallographic direction, e.g. (004). Additionally, we estimate the size of the characteristic nanocrystalline domains in carbon black from Scherrer's formula: the characteristic length parallel to the graphite basal plane, $L_a$, is 4.1 nm, and the characteristic length perpendicular to the graphite basal plane, $L_c$, is 1.7 nm. This result is consistent with the TEM micrograph displayed in Figure 2B as well as literature results of similar carbon blacks.[61,62] Importantly, we note XRD studies have shown that carbon blacks "differ only in the magnitude of their variation from graphite rather than representing different crystallographic structures."[62]

From pair distribution function analysis (Figure 2E), all carbon-carbon bond lengths in carbon black share indices with those of a single graphene layer. This result indicates carbon black and graphite have identical in-plane structures. Lastly, from C1s XPS spectra of pristine carbon black and graphite powders (Figure 2F), we find the carbon binding environment of the carbon black surface has similar character to graphite ($sp^2$).

Figure 3 displays electrochemical characterization of carbon black. In Figure 3a, voltage vs capacity are displayed during lithiation and delithiation for selected cycles at low rate (C/10). The absence of plateaus in the voltage curves beyond the first cycle indicates a suppression of graphitic phase separation,[70–72] as previously confirmed by in-situ XRD.[70] Solid-solution lithiation and lower reversible capacities are both consequences of disorder in nanostructured carbon.[63,64,70–75] Overall, the capacities are slightly higher than the values of 180–200 mAh g$^{-1}$ previously reported in literature.[35,48,68] We also note the large voltage plateau during the first lithiation at ~0.9 V, which we attribute to ethylene gas evolution from the reduction of ethylene carbonate.[56–59] The first cycle's unknown effect on subsequent cycling motivates our choice to standardize our formation cycling to be C/10 for all cells in this work and is the subject of a future investigation. As previously discussed, the initial formation of SEI on pristine carbon only occurs on the first lithiation so is thus trivially "asymmetric" (i.e., relative to the first delithiation), but this work focuses on directional asymmetry after the first cycle.

Due to their high specific surface area, carbon nanomaterials such as carbon black typically have high specific capacitance.[76] The linear, shallow-sloped region from ~0.5 V to ~1.0 V vs the Li counter/pseudo-reference electrode indicates a voltage regime with primarily, and significant, capacitive charge storage. The linearity of this region over a large potential range (0.6 V) suggests the capacitive capacity is a weak function of potential. The slope of a linear fit from 1.0 V to 0.5 V is 118 mAh g$^{-1}$ V$^{-1}$ in the second delithiation, equivalent to a specific capacitive capacity of 140 mAh g$^{-1}$ and a specific capacitance of 423 F g$^{-1}$ over the entire voltage range of 1.2 V to 0.01 V. This capacitive capacity is 63.3% of the total capacity. Multiple interfaces, such as carbon/electrolyte and SEI/electrolyte, can contribute to the total capacitance. Both intercalation and capacitance contribute significantly to charge storage in carbon black.

Figure 3b displays capacity vs cycle number for the same cell displayed in Figure 3a. The lithiation capacities decay sharply, while the delithiation capacities are relatively stable. This trend is nearly identical to the limiting case of completely asymmetric SEI growth on carbon lithiation (Figure 1a). This result is surprising, as we initially expect directional symmetric growth due to the similar voltage profiles in both directions. The use of a . This result provides an early indication of directionally asymmetric SEI growth.

Figures 3c and 3d plot the rate capability of carbon black during lithiation and delithiation. Figure 3c presents voltage vs specific capacity as a function of rate, while Figure 3d presents specific capacity vs cycle number during the rate test. In these experiments, a rate test was performed on two different carbon black cells after extensive cycling, which minimizes the contributions of SEI to the capacity. To isolate the rate performance of the direction under investigation, the cell was delithiated at C/10 for the lithiation rate test and lithiated at C/10 for the delithiation rate test. Carbon black retains 74.9% of its C/10 capacity at 5C during lithiation and 99.0% during delithiation. The rate capability is high on lithiation and even higher on delithiation; the large capacitance of carbon black provides a lower bound on the rate capability in both directions. Graphite electrodes also exhibit lower rate capability on lithiation; previous work typically reports ~10% of nominal capacity during lithiation at 5C.[77–79] For both materials, the lower rate capability on lithiation can be attributed to the cutoff voltage limitation imposed by the lithium plating potential, which reduces the maximum overpotential available at high rates. We note that C/5 appears to be the upper limit beyond which the capacity retention of carbon black drops non-negligibly on lithiation; carbon black retains 98.8% of its C/10 capacity at C/5 during lithiation. Thus, we assume that carbon black charge storage is in quasi-equilibrium, i.e., carbon black follows a similar voltage curve during lithiation and delithiation, when cycled at a rate below C/5.

***SEI growth measurement method.–*** Our analysis method relies on the SEI's self-passivating nature. SEI growth is high in early cycles and decreases with cycle number, eventually approaching zero growth. In principle, if we subtract the capacity in an early cycle, with contributions from both carbon black and SEI, from the capacity at a later "baseline" cycle, which primarily measures carbon black, we can attribute the difference to SEI.

Figure 4 illustrates this method of isolating SEI growth from other contributions to the measured capacity. Figure 4a displays capacity vs cycle number for a representative carbon black/Li half cell cycled at C/20 between 0.01 V and 2.0 V. Here, the "baseline" cycle is reached at ~$n=50$ for both directions. We note that the delithiation capacities are no longer constant with respect to cycle number as in Figure 3b, where the upper cutoff voltage was 1.2 V. Nevertheless, this capacity variation with cycle number is small compared to that of lithiation. $\Delta dQ/dV$ analysis, discussed in the following paragraphs, reveals that the small capacity change on delithiation primarily occurs between 1.2 V and 2.0 V, a potential regime in which we do not expect SEI growth. This change can be observed as small peaks centered at 1.3 V and 1.8 V in Figure 4e. This anomalous change in capacity could be due to reduction of electrolyte contaminants, dissolution of the copper current collector[80–82], or even electrochemically reversible SEI[48–55]. In any case, SEI growth is largely irreversible in the potential window studied here.

In Figures 4b-4e, we apply this principle to differential capacity, $dQ/dV$, to assess the voltage dependence of SEI growth. In contrast to integrative quantities like total capacity, $dQ/dV$ offers a continuous measurement of voltage dependence within a cycle. Previous work measured voltage dependence by discretizing the voltage region of interest over many cells.[83,84] In this work, we investigate the second cycle since the SEI is growing rapidly in this time domain. We plot *dQ/dV* of both cycle 2 and the baseline cycle for carbon lithiation (4b) and delithiation (4d). Under our current convention, *dQ/dV* is negative for lithiation (half cell discharging) and positive for delithiation (half cell charging). The gray shaded area represents the difference, or Δ*dQ/dV*, between cycle 2 and the baseline cycle, as plotted in Figures 4c and 4e. Capacity loss presents as negative values of Δ*dQ/dV* during both lithiation and delithiation, which we primarily attribute to SEI growth.

A key assumption in this analysis is that the carbon charge storage dynamics do not change with cycling. In other words, we assume quasistatic and invariant lithiation and delithiation of carbon black for every cycle. First, we consider rate-independent effects on the carbon capacity, primarily active material loss. Active material loss, in which individual electrode particles lose electronic or ionic connection to the remainder of the electrode, is a commonly reported degradation mode for graphitic negative electrodes in commercial lithium-ion cells.[30,85–87] However, if active material loss were appreciably present in our system, we would expect a decrease in delithiation capacity with respect to cycle number. In Figure 3b, we observe that the delithiation capacities are constant with respect to cycle number. Furthermore, we would also expect a decrease in differential capacity at all potentials, not just low potentials where carbon black stores more charge. We observe no changes in capacity at moderate potentials, i.e., Δ*dQ/dV* in Figure 5 is essentially zero from 0.5 V–1.0 V. Both arguments suggest that active material loss is minimal in our system.

We also consider rate-dependent effects, such as impedance increase and varying carbon black current. Impedance growth on either the carbon working electrode or the Li metal counter electrode will limit the reversible capacity that can be inserted or extracted from carbon black near the cutoff potentials. However, we expect impedance increase on both electrodes to be a small contributor at the low nominal rates we have selected. The highest rate studied here, C/5, corresponds to an absolute current of ~0.04 mA and an areal current of ~0.03 mA/cm$^2_{Li}$; 1.0 mA/cm$^2_{Li}$ is a typical areal current for cycling Li metal[88,89]. Lastly, as SEI growth decreases with cycle number, the current to carbon black will increase accordingly during galvanostatic (current-constrained) cycling, meaning the kinetics of carbon intercalation and bulk lithium transport may change with cycle number. However, the rate capability of carbon black changes only slightly below C/5 (Figures 3c and 3d), meaning the intercalation dynamics are stable at these low rates and the carbon electrode consistently reaches full lithiation (i.e., the carbon composition is consistent from cycle to cycle). Furthermore, the kinetics of capacitive charge storage is rapid. In summary, for a system without active material loss, we can effectively isolate SEI growth by cycling at low nominal rates.

We discuss three additional considerations here. Importantly, our technique does not measure SEI growth on the lithium metal counter electrode. While Li metal certainly can grow SEI, the Li metal capacity is much larger than that of carbon. As a result, its potential remains essentially constant with cycling, even when considering the redox potentials of SEI reactions. Thus, the destination of the lithium ions and electrons within the Li foil is irrelevant, and only SEI growth of the carbon black working electrode is measured. We also assume the impact of electrode "cross-talk", or interaction between SEI products of the negative and positive electrodes[90–96], between carbon and lithium is small, given the similarity of their SEI products[6,14] and the absence of transition metal dissolution from a transition metal oxide cathode. Lastly, we note that the baseline cycles are selected manually for each cell, meaning our result may be dependent on this choice. The cycle number of the baseline cycle varies with C rate, ranging from ~$n$=10 for C/100 to ~$n$=150 for C/5. However, we find the results are relatively insensitive to the baseline cycle chosen because the capacities near the baseline cycle are stable (e.g. see cycles 40-50 in Figure 4a).

***SEI growth as a function of C rate.*** – Figure 5 displays $\Delta dQ/dV$ between cycle 2 and the baseline cycle for five nominal rates ranging from C/100 to C/5. We select C/5 as our maximum nominal C rate to avoid rates at which the intercalation overpotential is significant, per Figure 3d. Thus, the intercalation process does not have significant kinetic limitations and can be assumed to be in quasi-equilibrium, and capacitive charge storage presumably occurs on even shorter timescales. In contrast, SEI growth is far from equilibrium as the reaction overpotential is high for most of the time per cycle. Because the first cycle was standardized at C/10 for all cells, the initial conditions at the start of the second lithiation are consistent. The three cells tested per C rate agree well, illustrating the high reproducibility of our cell fabrication and measurement. $\Delta dQ/dV$ of lithiation decreases with increasing nominal C rate, while the delithiation $\Delta dQ/dV$ is small and relatively invariant with nominal C rate. The asymmetry in SEI growth between carbon black lithiation and delithiation is stark for all measured conditions. Additionally, the change in SEI growth rate from high growth at the end of lithiation to low growth at the start of delithiation is essentially instantaneous.

$\Delta dQ/dV$ provides a continuous measurement of the voltage dependence of SEI growth that requires only a single cell, removing errors introduced by cell-to-cell variation (which can be

significant[97–99]). For all conditions, SEI growth becomes noticeable just below ~0.3 V and rapidly accelerates beyond ~0.1 V. This trend suggests a mechanism of SEI growth where the rate limiting step is expected to have strong voltage dependence. Given the thin SEI expected at this stage, SEI growth may still be reaction limited; the exponential dependence of overpotential on reaction rate given by the Butler-Volmer equation may account for the strong voltage dependence observed in Figure 5. The voltage regime of SEI growth is also similar to the voltage regime of lithium intercalation into carbon black, which may suggest these processes are related. Other properties of the SEI, such as electron transport ability[100] and chemistry/morphology[12], can exhibit strong voltage dependence and may also influence the observed voltage dependence of SEI growth. Another theory of voltage dependence is presented in Part II of this work.

Figure 6a summarizes the results of Figure 5 by subtracting the baseline capacity from the second cycle capacity from 0.01 V to 0.7 V. This differenced capacity is the total second-cycle SEI capacity for both lithiation and delithiation as a function of nominal C rate and is equivalent to integrating the second-cycle $\Delta dQ/dV$ with respect to voltage within this voltage range, with minimal noise. The strong dependence of SEI growth on C rate is evident for lithiation, as the SEI capacity decreases with C rate. Since the time per cycle varies with nominal C rate, we can divide the SEI capacity per cycle by the time per cycle to obtain an average SEI growth rate (in units of mA g$^{-1}$), as shown in Figure 6b. Interestingly, we now find that the average second-cycle SEI growth rate during lithiation increases roughly linearly with nominal C rate; the dimensionless slope of a linear fit of specific average SEI growth rate during lithiation (mA g$^{-1}$) to specific applied current (also mA g$^{-1}$) is 0.11 ± 0.01 (95% confidence interval). This result implies SEI growth consistently consumes ~11% of the total applied current during the second lithiation. This rate dependence of SEI growth contrasts with previous reports on graphite[26] as well as models that assume simple time-dependent growth[39–43], as these models would predict that the SEI growth rate is independent of the global C rate. In other words, the second-cycle SEI capacity would be linear with nominal C rate, and the second-cycle SEI growth rate (i.e., current) would be constant with nominal C rate. This result suggests SEI growth and charge storage in carbon black (intercalation and capacitance) are coupled, meaning the carbon electrode is more than just a substrate for SEI growth; in short, SEI growth and charge storage are not independent events.

As an aside, the dependence of SEI growth on C rate may have interesting implications for lithium-ion batteries operated at high charging rates, which is a major focus of research efforts in the field. While the dependence of SEI growth rate on C rate may be smaller for typical electrode materials such as graphite, SEI growth would be a major degradation mode during fast charging if the trend seen in Figure 6b continues at higher nominal C rates. Although we avoid rates exceeding C/5 in our experiments, we explore higher nominal C rates via modeling in Part II.

Summarizing our results so far, we have measured the dependencies of SEI growth on potential, current, and current direction using $\Delta dQ/dV$. First, we observe that SEI growth accelerates at low potential, which is consistent with literature reports[45,46]. Second, we find SEI growth rates strongly depend on the nominal C rate, at least for lithiation, which is unexpected given time-dependent models of SEI growth[39–43]. Lastly, we find that SEI growth is significantly higher on lithiation than delithiation. This finding is also unexpected, given the nearly identical voltage profiles during lithiation and delithiation at low rates, and incompatible with commonly accepted transport-limited models of SEI growth that only rely on time and potential. We explore the cause of these findings, particularly the directional asymmetry, further in the rest of this paper.

***Discussion of asymmetry in current direction.–*** We consider possible explanations for the observed asymmetry in current direction. Again, this asymmetry specifically refers to the dependence of SEI growth rate on the direction of carbon (de)intercalation at a given potential. In fact, our choice to perform galvanostatic cycling leads to a subtle source of directional dependence in our system. Our system has two primary destinations for current: the carbon black electrode and SEI growth. Regardless of the control conditions, the total current is given by $i_{total} = i_{CB} + i_{SEI}$, equivalent to a circuit with the carbon black and SEI components in parallel. During galvanostatic cycling, $i_{total}$ is constrained.

Again, $i_{total}$ is negative during lithiation (discharging of the half cell) and positive during delithiation (charging of the half cell). However, $i_{SEI}$ is always negative since the overpotential for SEI growth is negative (cathodic) for most of the lithiation and delithiation steps. Thus, on lithiation, $i_{total}$ and $i_{SEI}$ are both negative, meaning $i_{CB}$ is a smaller negative number than $i_{total}$. Because the SEI reduces the total current available for carbon black, carbon black lithiates at a low rate relative to the applied current. This effect increases the time spent in lithiation, allowing for more SEI to grow. During delithiation, $i_{SEI}$ remains negative while $i_{total}$ is positive, meaning $i_{CB}$ is larger positive number than $i_{total}$. Thus, the SEI current forces the carbon black current to be larger than the applied current, which reduces the time spent in delithiation. This effect shortens the time for SEI to grow, leading to lower overall SEI capacities. In summary, SEI growth during galvanostatic cycling will increase the lithiation time and decrease the delithiation time, leading to high SEI capacities on lithiation and low SEI capacities on delithiation.

We then consider the role of this effect in our system. From Figure 6b, the SEI capacity is around 11% of the baseline capacity during the second lithiation cycle, meaning $i_{SEI}$, and thus this "constant current asymmetry", is considerable for carbon black. However, this source of directional asymmetry would affect SEI growth during both lithiation and delithiation, meaning neither capacity is constant with cycle number as the SEI self-passivates. In Figures 3b, 5, and 6, we demonstrate that the SEI growth occurs primarily during lithiation and not during delithiation, meaning SEI grows asymmetrically on lithiation (Figure 1a). This asymmetry mode is thus insufficient to explain the extreme degree of asymmetry observed: another effect must be present, specifically to explain the near-zero growth on delithiation.

As an aside, we mention that constant current asymmetry will play a small role for electrode materials with micron-scale particle sizes, such as graphite, but is significant for nanostructured electrodes such as silicon nanoparticles.[101–103] This effect causes slower charging and faster discharging during galvanostatic cycling for any system with non-negligible SEI growth, which may be a feature or a challenge depending on the application.

For additional confirmation, we perform a potentiostatic experiment to confirm this directional asymmetry, as displayed in Figure 7. In controlled-potential settings, the expression $i_{total} = i_{CB} + i_{SEI}$ still applies, but the constraint on $i_{total}$ is removed; thus, this setting provides additional confirmation that our observed asymmetry is not solely a consequence of the control conditions. We perform standard cycling at C/20 between 0.01 V and 1.2 V, but we perform a five-hour constant-voltage hold at 0.1 V in both directions (Figure 7a). We then compare the current decays during the hold on both lithiation (Figure 7b) and delithiation (7c) as a function of cycle number. Immediately, we see the progression of SEI current decays is larger for lithiation than for delithiation. Then, by applying a similar method as before, we extract the SEI current by subtracting the current decay in the cycle 2 from that of the baseline cycle (Figure 7d), again assuming the intercalation dynamics of carbon are constant with cycle number. Here, we find the directional asymmetry persists even in a potentiostatic setting. Furthermore, we find $\Delta I$ of lithiation for both cycles 2 and 3 is larger than $\Delta I$ of delithiation in cycle 2, confirming that the directional asymmetry is not merely due to SEI forming first on lithiation. We conclude that the observed directional asymmetry is not just an outcome of our galvanostatic control conditions but has a deeper physical origin. This directional asymmetry in SEI growth is especially striking given the similarity of both intercalative and capacitive charge storage as a function of current direction at the rates used in this work.

*Hypotheses for directional asymmetry.–* Figure 8 illustrates three hypotheses that could explain both the extreme degree of asymmetry observed and the nonzero current dependence (Figure 6b). All hypotheses describe a coupling mechanism between SEI growth and charge storage modes in carbon black. This mechanism must account for the nearly instantaneous change in SEI growth rate from high to low at the end of lithiation to the start of delithiation, precluding chemical or morphological transformations of the SEI (which would have longer timescales). Additionally, the low nominal rates used in this work preclude mechanisms that rely on significant compositional heterogeneity within the electrode, as charge storage of carbon is in quasi-equilibrium.

First, we present a mechanical argument (Figure 8a). Carbon black and graphite particles expand by ~6% ([70]) and ~10% ([104,105]), respectively, during lithiation. SEI cracking has been observed during particle expansion, especially for materials with large volume expansion such as silicon.[49,106,107] If SEI can only grow on freshly exposed carbon surface, SEI growth would consequently only occur during lithiation. The comparatively small volume expansion of carbon, as well as the sudden change from high SEI growth at the end of carbon lithiation to low SEI growth at the beginning of carbon delithiation (Figure 5), indicates this mechanism is unlikely but is the subject of a future investigation.

A second source of directional asymmetry could arise simply from the intrinsic differences between lithiation and delithiation into carbon (Figure 8b), as well as the differences between capacitive storage and removal on the carbon surface. For example, their reaction rates (i.e., Tafel slopes) may be quite different. Thus, a directional difference in the (de)intercalation reaction rate could couple to a directional difference in SEI growth rate, leading to both the nonzero current dependence and the observed asymmetry in SEI growth. The reaction rate asymmetry can be captured phenomenologically by the deviation of the charge-transfer coefficient from the symmetric value of 0.5. Experimental measurements have measured charge-transfer coefficients of ~0.65 for graphite electrodes in electrolytes similar to those used in this study.[108,109] This charge-transfer asymmetry may be attributed to the high activation energy barrier of desolvation[110–115], which slows ion insertion (lithiation). In the context of asymmetric Marcus-Hush kinetics of electron transfer, charge-transfer asymmetry refers to a difference in the solvent reorganization free-energy curvatures for the reduced and oxidized states.[116–118] We study the effects of asymmetric charge transfer on outer SEI growth in the modeling effort presented in Part II. More generally, the molecular pathways (mechanisms) of lithiation and delithiation may be quite different. In other words, the intermediates of the lithiation and delithiation processes may be different and may selectively react to form SEI. Thus, the mechanism of SEI growth may simply be coupled specifically to the lithiation mechanism.

Third, we propose a novel source of directional asymmetry in SEI growth by considering the SEI as a nonideal mixed ionic-electronic conductor (MIEC), depicted in Figure 8c. For an intrinsic MIEC, the ionic and electronic concentrations are approximately equal. Thus, the ionic concentration controls the electron concentration, which in turn controls the electronic conductivity. Specifically, if SEI growth occurs at the electrolyte/SEI interface (electron-limited growth), the lithium ion concentration in the SEI could affect its electronic conductivity and thus the rate of SEI growth. By coupling an ohmic potential drop across the SEI to the surface potential, the concentration of lithium ions within the SEI would depend strongly on the direction of current. As a result, the electronic conductivity of the SEI would be high during lithiation and low during delithiation, which would lead to the observed directional asymmetry. This electronic conductivity decreases with increasing SEI thickness, so the SEI transitions to a pure ionic conductor instead of an MIEC. We model this hypothesis in depth in Part II of this work. Briefly, the results indicate that the directional asymmetry is more sensitive to electron conduction parameters than SEI reaction parameters, and electronic conductivity varies approximately with the square of the local lithium concentration.

## Conclusions

In this work, we establish the voltage, C rate, and current direction dependencies for SEI growth on carbon black. We identify carbon black as a good system for studies of SEI growth on carbon due to its high SEI growth rate and solid-solution lithiation pathway. We then present a method to isolate the electrochemical signature of SEI growth from reversible charge storage. Next, we reexamine the relationship of the second-cycle SEI growth rate to voltage and current magnitude/direction and find strong dependencies for each. Most notably, we find that SEI growth rates exhibit a roughly linear dependence on the global C rate, and that SEI growth occurs nearly exclusively during carbon lithiation. While the constraint on the total current imposed galvanostatic cycling is one source of this directional asymmetry, another effect is required to explain our observations. We present three hypotheses that consider a coupling between charge storage (e.g., intercalation and capacitance) and SEI growth, and we develop experimentally-validated physical models to explore these hypotheses in depth in Part II.

This work has implications for efforts to minimize battery degradation. We can apply this technique to other electrodes and electrolytes of interest to quantify SEI growth rates for different applications (i.e., cycling conditions), enabling improvements in materials development and additive selection. Furthermore, our results suggest that SEI growth is a major degradation mode during fast charging and that SEI growth is more significant during cycling than storage. Incorporating these insights, as well as the dependencies of voltage, current magnitude, and current direction, will significantly improve battery management system algorithms that rely on models using only time or capacity throughput. Finally, this work furthers our fundamental understanding of the elusive nature of SEI growth. Future work applying this method to graphite electrodes is ongoing.

**Acknowledgements**

We thank Prof. Chris Chidsey, Matthijs van den Berg, Kipil Lim, Iwnetim Abate, Norman Jin, Dr. Tyler Mefford, Geoff McConohy, Dr. Chia-Chin Chen, Dr. Yiyang Li, Dr. Che-Ning Ye, and Peter Csernica for insightful discussions. This work is sponsored by the Ford-Stanford Alliance. P.M.A. is supported by the Thomas V. Jones Stanford Graduate Fellowship and the National Science Foundation Graduate Research Fellowship under Grant No. DGE-114747. S.D. is supported by the Toyota Research Institute through D3BATT: Center for Data-Driven Design of Li-Ion Batteries. S.J.H. is supported by the Assistant Secretary for Energy Efficiency, Vehicle Technologies Office of the U.S. Department of Energy (U.S. DOE) under the Advanced Battery Materials Research (BMR) Program. This research used resources of the Advanced Photon

Source, a U.S. Department of Energy (DOE) Office of Science User Facility operated for the DOE Office of Science by Argonne National Laboratory under Contract No. DE-AC02-06CH11357. Part of this work was performed at the Stanford Nano Shared Facilities (SNSF), supported by the National Science Foundation under award ECCS-1542152. Any opinion, findings, and conclusions or recommendations expressed in this material are those of the authors(s) and do not necessarily reflect the views of the National Science Foundation.

# References

1. E. Peled, *J. Electrochem. Soc.*, **126**, 2047 (1979).

2. P. Arora, *J. Electrochem. Soc.*, **145**, 3647 (1998).

3. D. Aurbach, *J. Power Sources*, **89**, 206–218 (2000).

4. P. Verma, P. Maire, and P. Novák, *Electrochimica Acta*, **55**, 6332–6341 (2010).

5. V. Etacheri, R. Marom, R. Elazari, G. Salitra, and D. Aurbach, *Energy Environ. Sci.*, **4**, 3243 (2011).

6. E. Peled and S. Menkin, *J. Electrochem. Soc.*, **164**, A1703–A1719 (2017).

7. J. O. Besenhard, M. Winter, J. Yang, and W. Biberacher, *J. Power Sources*, **54**, 228–231 (1995).

8. Y. S. Cohen, Y. Cohen, and D. Aurbach, *J. Phys. Chem. B*, **104**, 12282–12291 (2000).

9. F. Kong et al., *J. Power Sources*, **97–98**, 58–66 (2001).

10. A. Xiao, L. Yang, B. L. Lucht, S.-H. Kang, and D. P. Abraham, *J. Electrochem. Soc.*, **156**, A318 (2009).

11. M. Nie et al., *J. Phys. Chem. C*, **117**, 1257–1267 (2013).

12. P. Lu, C. Li, E. W. Schneider, and S. J. Harris, *J. Phys. Chem. C*, **118**, 896–903 (2014).

13. R. Qiao et al., *Adv. Mater. Interfaces*, **1**, 1–6 (2014).

14. M. Gauthier et al., *J. Phys. Chem. Lett.*, **6**, 4653–4672 (2015).

15. A. L. Michan, M. Leskes, and C. P. Grey, *Chem. Mater.*, **28**, 385–398 (2016).

16. D. Aurbach, B. Markovsky, I. Weissman, E. Levi, and Y. Ein-Eli, *Electrochimica Acta*, **45**, 67–86 (1999).

17. K. Edström, M. Herstedt, and D. P. Abraham, *J. Power Sources*, **153**, 380–384 (2006).

18. K. W. Schroder, H. Celio, L. J. Webb, and K. J. Stevenson, *J. Phys. Chem. C*, **116**, 19737–19747 (2012).

19. R. R. Unocic et al., *Microsc. Microanal.*, **20**, 1029–1037 (2014).

20. F. Lin, I. M. Markus, M. M. Doeff, and H. L. Xin, *Sci. Rep.*, **4** (2015) http://www.nature.com/articles/srep05694.

21. Y. Li et al., *Science*, **358**, 506–510 (2017).

22. A. J. Smith, J. C. Burns, and J. R. Dahn, *Electrochem. Solid-State Lett.*, **13**, A177 (2010).

23. A. J. Smith, J. C. Burns, S. Trussler, and J. R. Dahn, *J. Electrochem. Soc.*, **157**, A196–A202 (2010).

24. A. J. Smith, J. C. Burns, D. Xiong, and J. R. Dahn, *J. Electrochem. Soc.*, **158**, A1136 (2011).

25. J. C. Burns et al., *J. Electrochem. Soc.*, **158**, A255 (2011).

26. A. J. Smith, J. C. Burns, X. Zhao, D. Xiong, and J. R. Dahn, *J. Electrochem. Soc.*, **158**, A447–A452 (2011).

27. I. Bloom et al., *J. Power Sources*, **139**, 295–303 (2005).

28. M. Dubarry and B. Y. Liaw, *J. Power Sources*, **194**, 541–549 (2009).

29. A. J. Smith, J. C. Burns, and J. R. Dahn, *Electrochem. Solid-State Lett.*, **14**, A39 (2011).

30. M. Safari and C. Delacourt, *J. Electrochem. Soc.*, **158**, A1123–A1135 (2011).

31. A. J. Smith and J. R. Dahn, *J. Electrochem. Soc.*, **159**, A290 (2012).

32. G. E. Blomgren, 7 (1999).

33. M. Broussely et al., *J. Power Sources*, 9 (2001).

34. M. Broussely et al., *J. Power Sources*, **146**, 90–96 (2005).

35. A. J. Smith, J. C. Burns, X. Zhao, D. Xiong, and J. R. Dahn, *J. Electrochem. Soc.*, **158**, A447 (2011).

36. A. J. Smith, J. C. Burns, D. Xiong, and J. R. Dahn, *J. Electrochem. Soc.*, **158**, A1136 (2011).

37. I. Bloom et al., *J. Power Sources*, **101**, 238–247 (2001).

38. R. . Wright et al., *J. Power Sources*, **119–121**, 865–869 (2003).

39. J. Christensen and J. Newman, *J. Electrochem. Soc.*, **151**, A1977 (2004).

40. M. Tang, S. Lu, and J. Newman, *J. Electrochem. Soc.*, **159**, A1775–A1785 (2012).

41. M. Broussely et al., *J. Power Sources*, **97–98**, 13–21 (2001).

42. H. J. Ploehn, P. Ramadass, and R. E. White, *J. Electrochem. Soc.*, **151**, A456 (2004).

43. M. B. Pinson and M. Z. Bazant, *J. Electrochem. Soc.*, **160**, A243–A250 (2012).

44. F. Single, B. Horstmann, and A. Latz, *J. Electrochem. Soc.*, **164**, E3132–E3145 (2017).

45. P. Keil et al., *J. Electrochem. Soc.*, **163**, 1872–1880 (2016).

46. J. Xia, M. Nie, L. Ma, and J. R. Dahn, *J. Power Sources*, **306**, 233–240 (2016).

47. B. E. Deal and A. S. Grove, *J. Appl. Phys.*, **36**, 3770–3778 (1965).

48. K. A. See, M. A. Lumley, G. D. Stucky, C. P. Grey, and R. Seshadri, *J. Electrochem. Soc.*, **164**, A327–A333 (2017).

49. C. K. Chan, R. Ruffo, S. S. Hong, and Y. Cui, *J. Power Sources*, **189**, 1132–1140 (2009).

50. K. Tasaki et al., *J. Electrochem. Soc.*, **156**, A1019 (2009).

51. K. Tasaki and S. J. Harris, *J. Phys. Chem. C*, **114**, 8076–8083 (2010).

52. S. J. Rezvani et al., *ACS Appl. Mater. Interfaces*, **9**, 4570–4576 (2017).

53. B. S. Parimalam, A. D. MacIntosh, R. Kadam, and B. L. Lucht, *J. Phys. Chem. C*, **121**, 22733–22738 (2017).

54. Z. Zhuo et al., *Chem. Commun.*, **54**, 814–817 (2018).

55. T. Liu et al., *Nat. Nanotechnol.* (2018) http://www.nature.com/articles/s41565-018-0284-y.

56. M. E. Spahr et al., *J. Electrochem. Soc.*, **151**, A1383 (2004).

57. M. E. Spahr et al., *J. Power Sources*, **153**, 300–311 (2006).

58. D. Goers, M. E. Spahr, A. Leone, W. Märkle, and P. Novák, *Electrochimica Acta*, **56**, 3799–3808 (2011).

59. R. Bernhard, M. Metzger, and H. A. Gasteiger, *J. Electrochem. Soc.*, **162**, A1984–A1989 (2015).

60. H. Bryngelsson, M. Stjerndahl, T. Gustafsson, and K. Edström, *J. Power Sources*, **174**, 970–975 (2007).

61. K. Kinoshita, *Carbon: Electrochemical and Physiochemical Properties*, 1st ed., John Wiley & Sons, Inc., (1988).

62. J.-B. Donnet, R. C. Bansal, and M.-J. Wang, Eds., *Carbon Black: Science and Technology*, 2nd ed., Marcel Dekker, New York, NY, USA, (1993).

63. M. Winter and J. O. Besenhard, *Adv. Mater.*, **10**, 725–763 (1998).

64. N. A. Kaskhedikar and J. Maier, *Adv. Mater.*, **21**, 2664–2680 (2009).

65. A. K. Sleigh and U. von Sacken, *Solid State Ion.*, **57**, 99–102 (1992).

66. K. Takei et al., *J. Power Sources*, **55**, 191–195 (1995).

67. R. Yazami and M. Deschamps, *J. Power Sources*, **54**, 411–415 (1995).

68. L. Fransson, T. Eriksson, K. Edström, T. Gustafsson, and J. O. Thomas, *J. Power Sources*, **101**, 1–9 (2001).

69. J. Syzdek, M. Marcinek, and R. Kostecki, *J. Power Sources*, **245**, 739–744 (2014).

70. J. R. Dahn, R. Fong, and M. J. Spoon, *Phys. Rev. B*, **42**, 6424–6432 (1990).

71. T. Zheng, J. N. Reimers, and J. R. Dahn, *Phys. Rev. B*, **51**, 734–741 (1995).

72. D. A. Stevens and J. R. Dahn, *J. Electrochem. Soc.*, **148**, A803 (2001).

73. J. R. Dahn et al., *Electrochimica Acta*, **38**, 1179–1191 (1993).

74. S. Flandrois and B. Simon, *Carbon*, **37**, 165–180 (1999).

75. Y. F. Reynier, R. Yazami, and B. Fultz, *J. Electrochem. Soc.*, **151**, A422 (2004).

76. M. R. Lukatskaya, B. Dunn, and Y. Gogotsi, *Nat. Commun.*, **7**, 12647 (2016).

77. H. Buqa, D. Goers, M. Holzapfel, M. E. Spahr, and P. Novák, *J. Electrochem. Soc.*, **152**, A474 (2005).

78. S. R. Sivakkumar, J. Y. Nerkar, and A. G. Pandolfo, *Electrochimica Acta*, **55**, 3330–3335 (2010).

79. J. Billaud, F. Bouville, T. Magrini, C. Villevieille, and A. R. Studart, *Nat. Energy*, **1**, 16097 (2016).

80. S. Grugeon et al., *J. Electrochem. Soc.*, **148**, A285 (2001).

81. R. Carter, B. Huhman, C. T. Love, and I. V. Zenyuk, *J. Power Sources*, **381**, 46–55 (2018).

82. C. Fear, D. Juarez-Robles, J. A. Jeevarajan, and P. P. Mukherjee, *J. Electrochem. Soc.*, **165**, A1639–A1647 (2018).

83. P. Keil et al., *J. Electrochem. Soc.*, **163**, A1872–A1880 (2016).

84. J. Xia, M. Nie, L. Ma, and J. R. Dahn, *J. Power Sources*, **306**, 233–240 (2016).

85. P. Liu et al., *J. Electrochem. Soc.*, **157**, A499–A507 (2010).

86. D. Anseán et al., *J. Power Sources*, **321**, 201–209 (2016).

87. M. Lewerenz, A. Marongiu, A. Warnecke, and D. U. Sauer, *J. Power Sources*, **368**, 57–67 (2017).

88. P. Albertus, S. Babinec, S. Litzelman, and A. Newman, *Nat. Energy*, **3**, 16–21 (2018).

89. F. Shi et al., *Proc. Natl. Acad. Sci.*, 201806878 (2018).

90. C. Delacourt et al., *J. Electrochem. Soc.*, **160**, A1099–A1107 (2013).

91. I. A. Shkrob et al., *J. Phys. Chem. C*, **118**, 24335–24348 (2014).

92. T. Joshi, K. Eom, G. Yushin, and T. F. Fuller, *J. Electrochem. Soc.*, **161**, A1915–A1921 (2014).

93. D. J. Xiong et al., *J. Electrochem. Soc.*, **163**, A3069–A3077 (2016).

94. J. A. Gilbert, I. A. Shkrob, and D. P. Abraham, *J. Electrochem. Soc.*, **164**, A389–A399 (2017).

95. B. Michalak, B. B. Berkes, H. Sommer, T. Brezesinski, and J. Janek, *J. Phys. Chem. C*, **121**, 211–216 (2017).

96. K. W. Knehr et al., *Joule*, **2**, 1146–1159 (2018).

97. T. Baumhöfer, M. Brühl, S. Rothgang, and D. U. Sauer, *J. Power Sources*, **247**, 332–338 (2014).

98. S. F. Schuster, M. J. Brand, P. Berg, M. Gleissenberger, and A. Jossen, *J. Power Sources*, **297**, 242–251 (2015).

99. S. J. Harris, D. J. Harris, and C. Li, *J. Power Sources*, **342**, 589–597 (2017).

100. S. Shi, Y. Qi, H. Li, and L. G. Hector, *J. Phys. Chem. C*, **117**, 8579–8593 (2013).

101. L. Y. Beaulieu, K. W. Eberman, R. L. Turner, L. J. Krause, and J. R. Dahn, *Electrochem. Solid-State Lett.*, **4**, A137 (2001).

102. M. N. Obrovac and L. Christensen, *Electrochem. Solid-State Lett.*, **7**, A93 (2004).

103. U. Kasavajjula, C. Wang, and A. J. Appleby, *J. Power Sources*, **163**, 1003–1039 (2007).

104. J. R. Dahn, *Phys. Rev. B*, **44**, 9170–9177 (1991).

105. S. Schweidler et al., *J. Phys. Chem. C*, **122**, 8829–8835 (2018).

106. S. P. V. Nadimpalli et al., *J. Power Sources*, **215**, 145–151 (2012).

107. A. Tokranov, B. W. Sheldon, C. Li, S. Minne, and X. Xiao, *ACS Appl. Mater. Interfaces*, **6**, 6672–6686 (2014).

108. T. Piao, *J. Electrochem. Soc.*, **146**, 2794 (1999).

109. Y.-C. Chang, J.-H. Jong, and G. T.-K. Fey, *J. Electrochem. Soc.*, **147**, 2033 (2000).

110. T. Abe, H. Fukuda, Y. Iriyama, and Z. Ogumi, *J. Electrochem. Soc.*, **151**, A1120 (2004).

111. K. Xu, *J. Electrochem. Soc.*, **154**, A162 (2007).

112. Y. Yamada, Y. Iriyama, T. Abe, and Z. Ogumi, *Langmuir*, **25**, 12766–12770 (2009).

113. K. Xu, A. von Cresce, and U. Lee, *Langmuir*, **26**, 11538–11543 (2010).

114. Y. Yamada et al., *J. Am. Chem. Soc.*, **136**, 5039–5046 (2014).

115. Q. Li et al., *ACS Appl. Mater. Interfaces*, **9**, 42761–42768 (2017).

116. R. A. Marcus, *J. Chem. Phys.*, **43**, 679–701 (1965).

117. M. C. Henstridge, E. Laborda, and R. G. Compton, *J. Electroanal. Chem.*, **674**, 90–96 (2012).

118. Y. Zeng, P. Bai, R. B. Smith, and M. Z. Bazant, *J. Electroanal. Chem.*, **748**, 52–57 (2015).

# Figure Captions

**Figure 1.** Schematics illustrating capacity vs cycle number for symmetric and asymmetric SEI growth as a function of current direction. The dotted red line represents the reversible capacity of the electrode. (a) Completely asymmetric SEI growth on lithiation, in which the capacities measured on lithiation are the sum of the reversible capacity of the electrode and the SEI capacity, which decays with cycle number due to self-passivation. (b) Completely symmetric SEI growth, in which the capacities measured on delithiation are the difference between the reversible electrode capacity and the decaying SEI capacity. (c) Completely asymmetric SEI growth on delithiation. The contributions to ionic current on lithiation (d) and delithiation (e) include both the reversible electrode and the SEI.

**Figure 2.** Physical characterization of TIMCAL Super P carbon black. (a) SEM micrograph of carbon black electrode. (b) TEM micrograph of carbon black particle. The graphitic nanodomains, approximately 2 nm in size, are visible. (c) XRD patterns of graphite and carbon black with indexed peaks. (d) Magnification of the interplanar spacing peak (002) in XRD. (e) Synchrotron PDF pattern of carbon black. The yellow dots represent carbon-carbon distances calculated for a graphene plane. (f) C1s XPS spectra for graphite and carbon black.

**Figure 3.** Electrochemical characterization of TIMCAL Super P carbon black. (a) Voltage vs capacity during lithiation (blue) and delithiation (red) for a carbon black/Li half cell nominally cycled at C/10 between 0.01 and 1.2 V. Note the anomalous plateau of the first lithiation. (b) Capacity vs cycle number during lithiation and delithiation for the same cell in (a). This plot

resembles the asymmetric growth on lithiation case presented in Figure 1a. The capacity of the first lithiation (not shown) is 536 mAh g$^{-1}$. (c) Voltage vs capacity of carbon black cells as a function of rate for lithiation (blue) and delithiation (red). (d) Capacity vs cycle number of carbon black cells as a function of rate during lithiation and delithiation. In both (c) and (d), the (de)lithiation step not under investigation is cycled at C/10 to isolate the rate of each step; for example, the lithiation rate test is performed with C/10 delithiation steps.

**Figure 4.** Method of measuring SEI growth via voltage-dependent coulometry. (a) Capacity vs cycle number for a carbon black/Li half cell nominally cycled at C/20 between 0.01 and 2.0 V. The baseline cycle is selected by determining the point at which the lithiation and delithiation capacities are stable with cycle number, as illustrated by the dashed and dotted lines, respectively. The capacity of the first lithiation (not shown) is 539 mAh g$^{-1}$. (b) *dQ/dV* of cycle 2 and the baseline cycle (cycle 50) during lithiation as a function of voltage. The grey region represents the difference as a function of voltage, or Δ*dQ/dV*, which is also displayed in (c). (d) *dQ/dV* of cycle 2 and the baseline cycle (cycle 50) during delithiation as a function of voltage. The grey region represents the difference as a function of voltage, or Δ*dQ/dV*, which is also displayed in (e). The colored arrows indicate the direction of the change in voltage with time. In these plots, *dQ/dV* is negative for lithiation and positive for delithiation; capacity loss presents as negative values of Δ*dQ/dV* during both lithiation and delithiation.

**Figure 5.** Δ*dQ/dV* measurements between cycle 2 and the baseline cycle for five different nominal C rates: (a) C/100, (b) C/50, (c) C/20, (d) C/10, and (e) C/5. The blue and red lines represent lithiation and delithiation, respectively. Three cells were measured per condition and are represented by solid, dashed, and dotted lines. The colored arrows indicate the direction of the change in voltage with time.

**Figure 6.** Second-cycle SEI growth as a function of current magnitude and current direction, averaged over three cells. (a) Dependence of second-cycle SEI growth on nominal C rate. These capacities are calculated from the integral of the $\Delta dQ/dV$ curves in Figure 5. (b) Dependence of second-cycle average SEI growth rate on nominal C rate. The average SEI growth rate is calculated by dividing SEI growth by the time per cycle. A linear fit of the lithiation data to nominal applied current yields the equation $y = (0.11 \pm 0.01)x + (0.5 \pm 0.3$ mA g$^{-1})$, where errors represent the 95% CI. In both subplots, the error bars represent 95% confidence intervals of the mean.

**Figure 7.** SEI growth during cycling with potentiostatic interruptions. (a) Illustration of potentiostatic experiment. The cell nominally cycles at C/20 with a 5 hour constant-voltage hold performed at 100 mV in both directions. (b, c) Current vs time for lithiation (c) and delithiation (d) as a function of cycle number. Cycle 13 was selected as the baseline cycle. (d) $\Delta I$ between cycles 2 and 13 as a function of time for lithiation and delithiation, as well as $\Delta I$ between cycles 3 and 13 for lithiation. The directional asymmetry in SEI growth between lithiation and delithiation persists in potentiostatic mode.

**Figure 8.** Schematics of hypotheses for the observed asymmetry in SEI growth as a function of current direction. (a) Carbon particles expand during lithiation and contract during delithiation. Particle expansion may cause cracking in the SEI if the SEI is mechanically fragile, while cracking is less likely on particle contraction. If SEI growth is proportional to the freshly exposed surface

area of carbon black, new SEI would only grow on lithiation. (b) The net (de)intercalation reaction, $x\text{Li}^+ + x\text{e}^- + \text{Li}_y\text{C}_6 \rightarrow \text{Li}_{x+y}\text{C}_6$, is reversible, but an asymmetry in either the rate (phenomenologically captured by the charge-transfer coefficient) or the mechanistic pathway could couple to the observed asymmetry in SEI growth. This mechanism could also apply to the capacitive charge storage process. (c) We consider the SEI to be an intrinsic mixed ionic electronic conductor, for which the ionic and electronic concentrations are approximately equal. Thus, the ionic concentration controls the electron concentration, which in turn controls the electronic conductivity. Specifically, for electron-limited SEI growth, the lithium ion concentration in the SEI could affect its electronic conductivity, and thus the rate of SEI growth. If the lithium ion concentration in the SEI was a function of current direction, we would achieve the observed directional asymmetry.

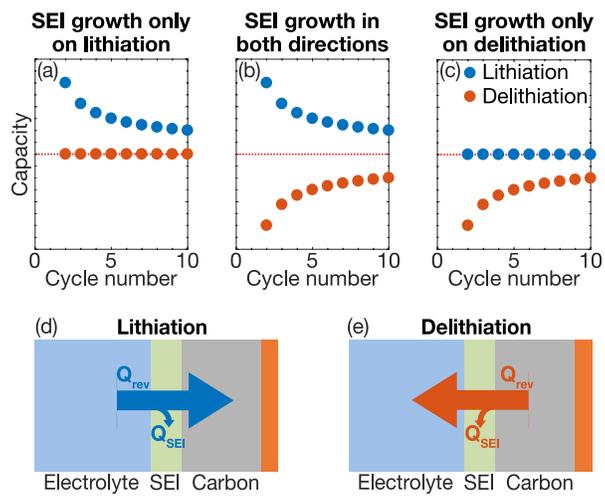

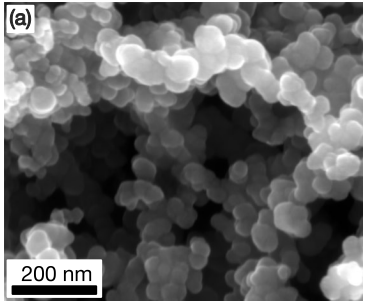
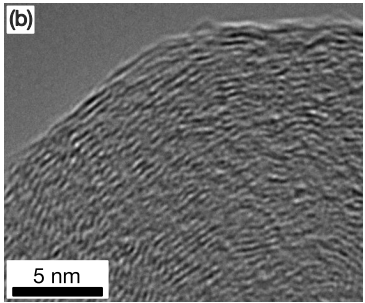
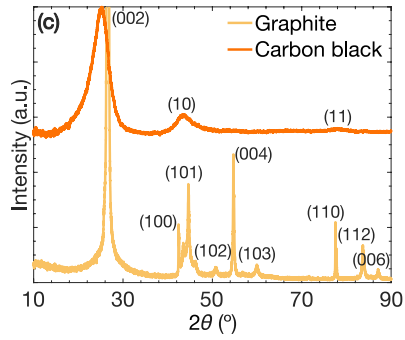
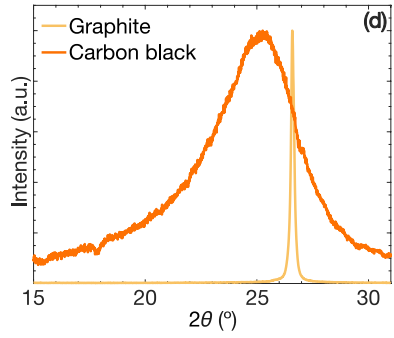
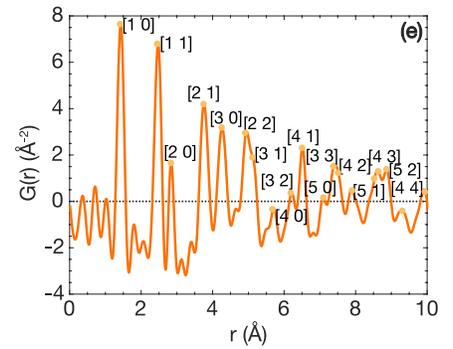
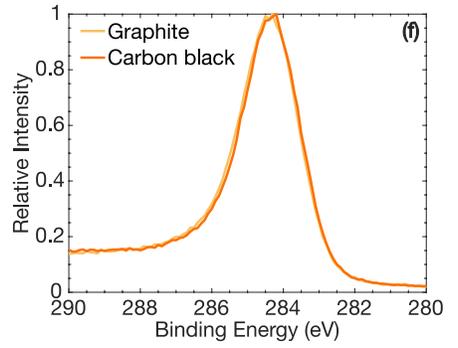

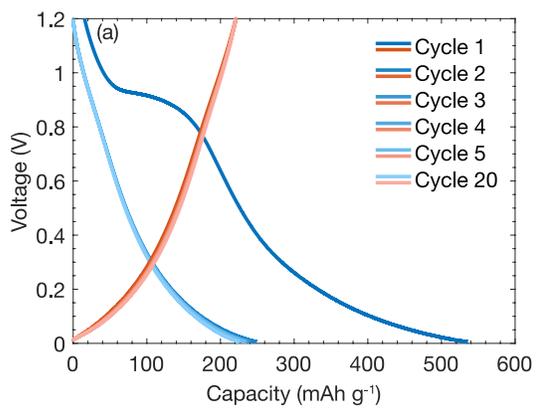
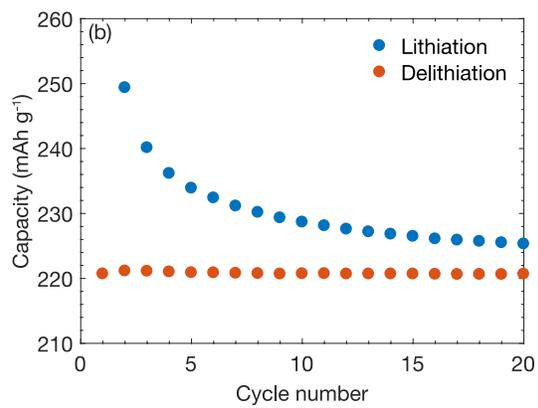
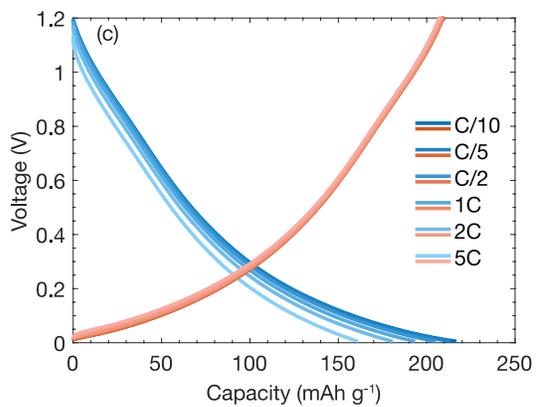
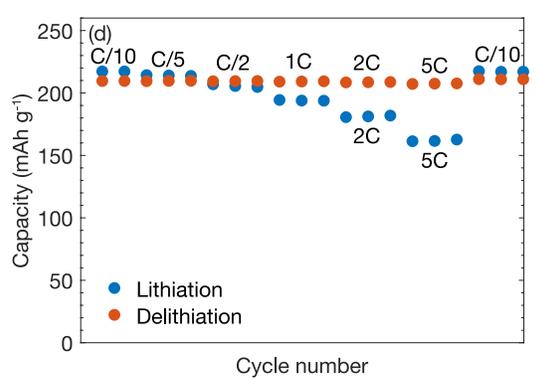

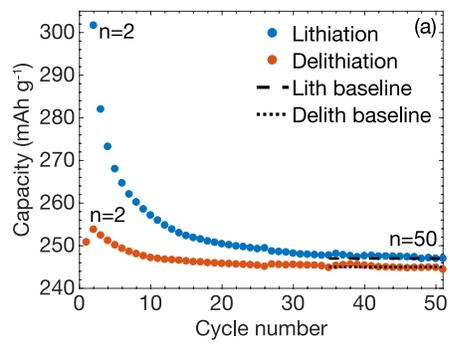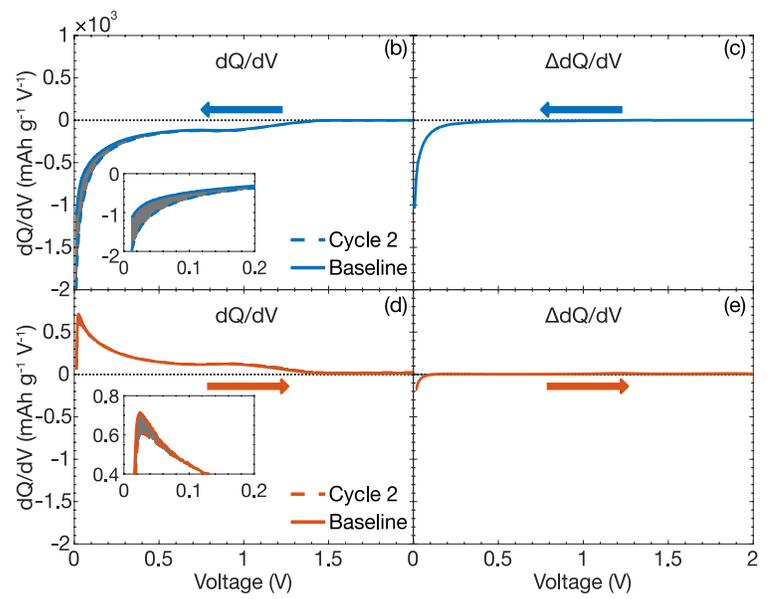

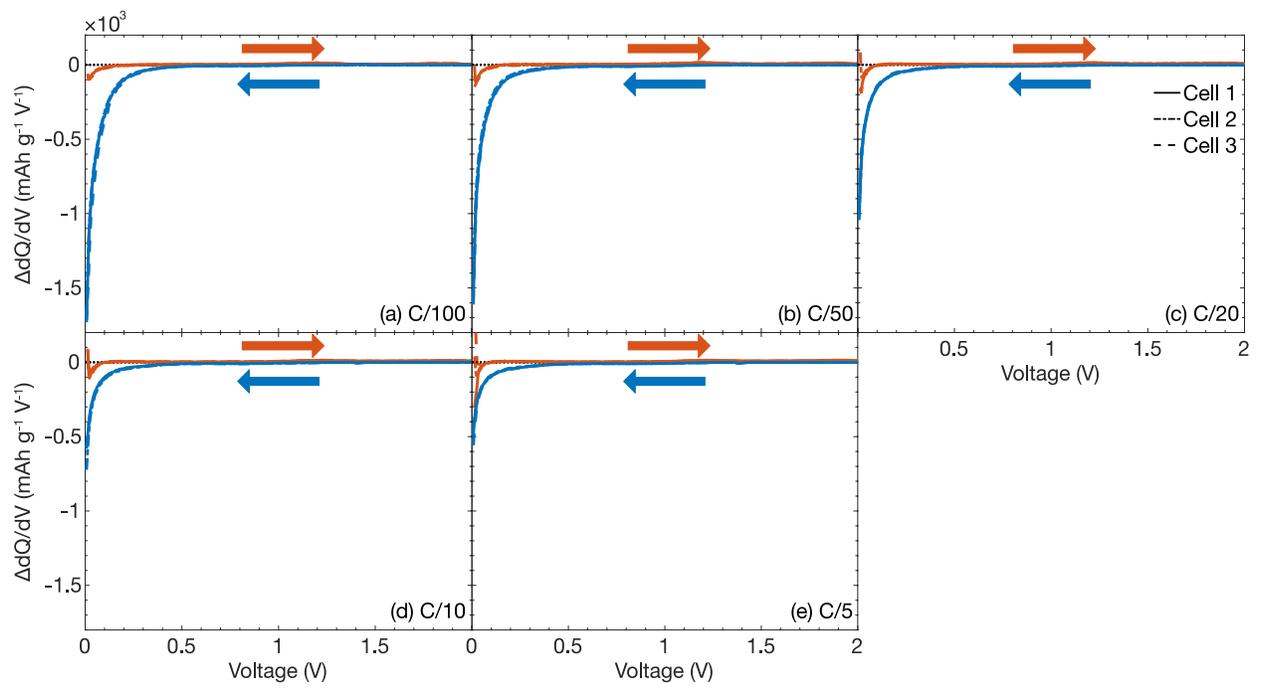

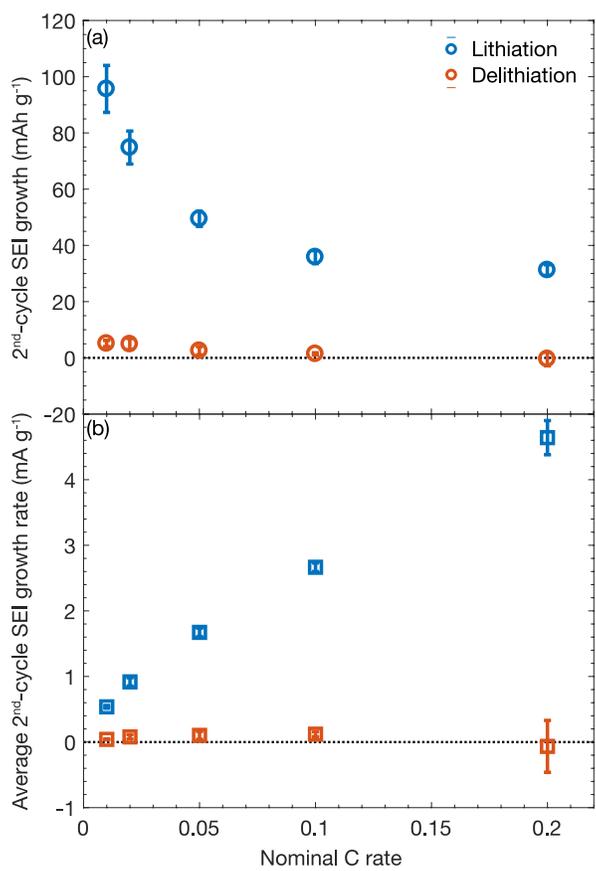

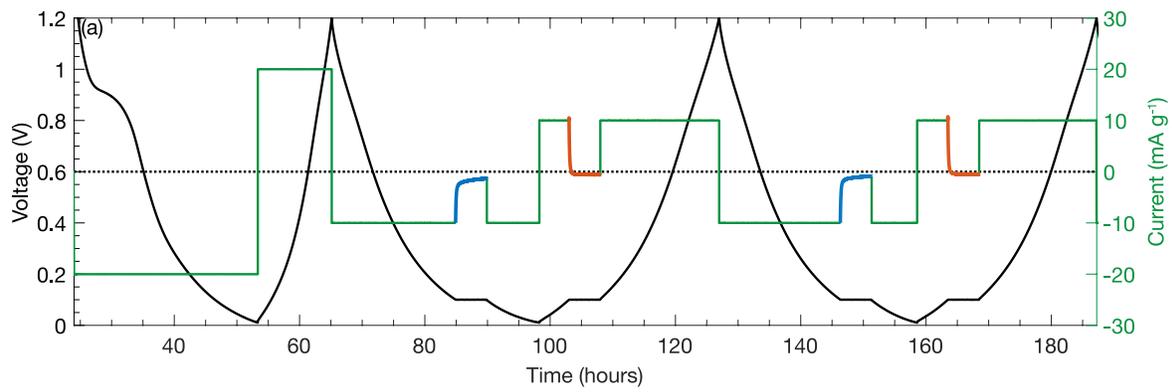
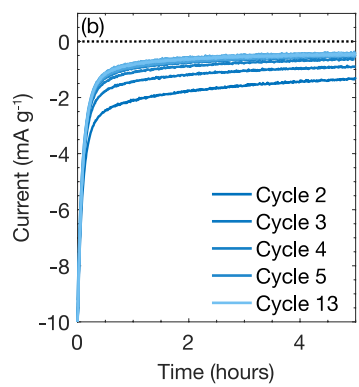
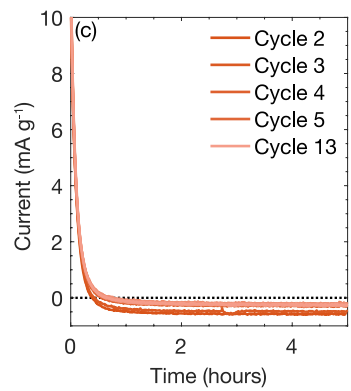
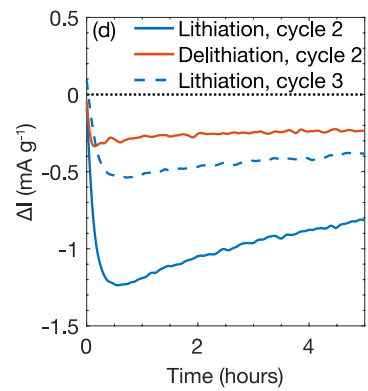

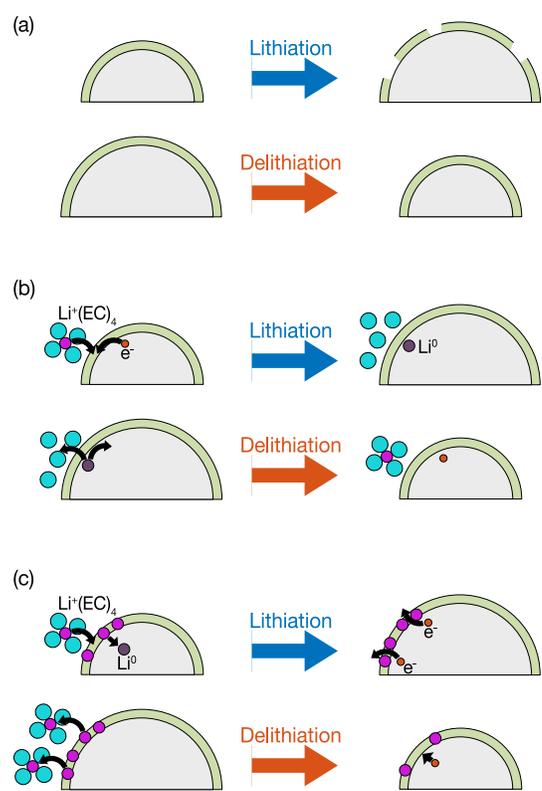